\documentclass{egs}                  

  \usepackage{times}                   

\usepackage{epsfig}
\usepackage{graphicx}

\printfigures              

\begin{document}
\title{Computing nonlinear force free coronal magnetic fields.}

%
%
%
%
\author[1,2]{T. Wiegelmann}
\affil[1]{School of Mathematics and Statistics,
University of St. Andrews, St. Andrews, KY16 9SS,
United Kingdom}
\author[1]{T. Neukirch}
\affil[2]{Max-Planck-Institut f\"ur Aeronomie,
Max-Planck-Strasse 2, 37191 Katlenburg-Lindau, Germany}

%
%

\journal{\NPG}       
%
%
\firstauthor{Wiegelmann}
\proofs{T. Wiegelmann}
\offsets{T. Wiegelmann}

\msnumber{Nonlinear Processes in Geophysics, Volume 10, Issue 4/5, 2003, pp.313-322}
\received{12. August 2002}
\revised{22. November 2002} 
\accepted{27. November 2002}

\runninghead{Coronal magnetic fields.}
\firstpage{1}
\pubyear{2002}
\pubvol{25}
\pubnum{2}

\maketitle

%
%
%

\begin{abstract}
Knowledge of the structure of the coronal magnetic field is
important for our understanding of many solar activity phenomena,
e.g. flares and CMEs. However, the direct measurement of coronal
magnetic fields
is not possible with present methods, and therefore the coronal field
has to be extrapolated from photospheric measurements.
Due to the low plasma beta the coronal magnetic field can usually be
assumed to be approximately force free, with electric currents
flowing along the magnetic field lines.
There are both observational and theoretical
reasons which suggest that at least
prior to an eruption the coronal
magnetic field is in a
nonlinear force free state. Unfortunately
the computation of nonlinear force free fields
is way more difficult than potential
or linear force free fields and analytic solutions are not generally available.
We discuss several methods which have been proposed
to compute nonlinear force free fields and focus particularly
on an optimization method which has been suggested recently.
We compare the numerical performance of a newly developed numerical code
based on the optimization method with the performance of
another code based on an MHD relaxation method if both codes are
applied to the reconstruction of a
semi-analytic nonlinear force-free solution.
The optimization method has also been tested for cases where we add random
noise to the perfect boundary conditions of the analytic solution, in this
way mimicking the more realistic case where the boundary conditions
are given by vector magnetogram data.
We find that the convergence properties of the optimization
method are affected by adding noise to the boundary data and we discuss
possibilities to overcome this difficulty.
\end{abstract}

\section{Introduction}
\label{sec:intro}

The magnetic field is the most important quantity for understanding the
plasma structure of the solar corona and the activity phenomena
it shows. Unfortunately a systematic direct measurement of the coronal
magnetic field is not feasible with the observational methods
presently available. Therefore the extrapolation of the magnetic field into
the corona from measurements of the magnetic field at the photospheric level is
extremely important.

It is generally assumed that the magnetic pressure in the corona is much
higher than the plasma pressure (small plasma $\beta$) and that
therefore the magnetic field is nearly force-free \citep[for a critical
view of this assumption see][]{gary01}. The extrapolation methods
based on this assumption include potential field extrapolation \citep[e.g.,][]{schmidt64,semel67},
linear force-free field extrapolation \citep[e.g.,][]{chiu:hilton77,seehafer78,seehafer82,semel88} and
nonlinear force-free field extrapolation \citep[e.g.,][]{amari:etal97}.
Methods for non-force-free
field extrapolation have also been developed \citep{petrie:neukirch00}
but are not used routinely.

Whereas potential and linear force-free fields can be used as a first step to
model the general structure of magnetic fields in the solar corona, the use of
non-linear force free fields is essential to understand eruptive phenomena as
there are both observational and theoretical reasons which suggest that
the pre-eruptive magnetic fields are
non-linear force-free fields (see below).
The calculation of nonlinear force free fields is
complicated by the intrinsic nonlinearity of the underlying mathematical
problem. Another problem which becomes especially important when nonlinear
force free field are used for magnetic field extrapolation is the
correct formulation of the problem with respect to boundary values.
The available magnetograph observation provide either the line-of-sight
magnetic field ($B_{los}$), which is sufficient for potential and
linear force free fields, or all three components of the photospheric
magnetic field. Only the latter information is sufficient to determine
non-linear force free fields completely. There are, however, tremendous
difficulties to be overcome both in extracting the necessary information
about the magnetic field components on the boundary from the data in an
unambiguous way and in the use of this information in the computation
of the corresponding magnetic field. Even though a lot of research has been
devoted to these problems in the past
so far none of the proposed methods has been found to be outstanding in combining
simplicity, usability and robustness.

The purpose of the present contribution is to assess some of the methods
presently available for calculating non-linear
force-free fields, with a view to implementing them into a general extrapolation code
which also takes stereoscopic information into account in the extrapolation
process.
A first version of such a method based on linear force free extrapolation
has recently been proposed \citep{twtn02}. As stated in \citet{twtn02} and
bearing in mind the remarks above, it is would be very desirable to generalize
the extrapolation method to nonlinear force free fields.

As a first step in this direction we investigate the robustness of the
optimization method
recently proposed by \citet{wheatland00}, and compare
its performance with another method based on MHD relaxation.
The paper is organized as follows.
In Sect. \ref{sec:basic} we discuss the basic equations and background theory.
Section \ref{sec:methods} describes various methods which have been proposed for
nonlinear force free field extrapolation. Out of these we focus
on the optimisation method and, for the case of
analytically given boundary data, on the MHD relaxation method.
These methods are tested
using an analytical solution in Sect. \ref{sec:tests}. The optimisation method is also
subjected to tests with more general boundary conditions where we add
noise to the given analytical boundary data.
The paper closes with the conclusions in Sect. \ref{sec:end}.

\section{Basic equations: Force-Free Equilibria}
\label{sec:basic}

For force free fields the equations to solve are
\begin{eqnarray}
{\bf j}\times{\bf B}  & = & {\bf 0}        \label{forcebal}\\
\nabla \times {\bf B }& = & \mu_0 {\bf j}  \label{ampere}, \\
\nabla\cdot{\bf B}    & = &         0      \label{solenoidal}.
\end{eqnarray}
Equation (\ref{forcebal}) implies that for force free fields the current density
and the magnetic field are parallel, i.e.
\begin{equation}
\mu_0 {\bf j} = \alpha {\bf B}.
\label{jparb}
\end{equation}
Here $\alpha$ is a function of space. This function has to satisfy the
equation
\begin{equation}
{\bf B} \cdot \nabla \alpha  = 0
\label{alphaeq}
\end{equation}
which is obtained by taking the divergence of Eq. (\ref{jparb}) and
making use of Eqs. (\ref{ampere}) and (\ref{solenoidal}). Substituting
Eq. (\ref{jparb}) into Eq. (\ref{ampere}) we get a second equation
which determines ${\bf B}$~:
\begin{equation}
\nabla \times {\bf B }  =  \alpha {\bf  B}
\label{amperealpha}
\end{equation}

Equation (\ref{alphaeq}) implies that $\alpha$ is constant
along magnetic field lines,
but that it can vary across the magnetic field. Obviously, Eq. (\ref{alphaeq})
is solved by $\alpha=0$ implying that ${\bf j} ={\bf 0}$ by Eq. (\ref{jparb}).
In this case we obtain potential fields. Another obvious possibility
is $\alpha$ constant, but nonzero. This is the case of linear (or
constant-$\alpha$) force free fields.
Standard methods for calculating
potential and linear
force free field are available
\citep[see e.g.][]{gary89}
and they can at least give a rough impression of the
coronal magnetic field
structure.

There are, however, both observational and theoretical arguments
that at least the magnetic field prior to eruptive processes in the corona
is not a linear force free (or potential) field. If the normal
component or the line-of-sight component of the magnetic
field on the boundary is given the corresponding potential
field is uniquely determined and is actually the magnetic field
with the lowest magnetic energy for this boundary condition. Since
magnetic energy is needed to drive an eruptive process pre-eruptive
coronal magnetic fields cannot be expected to be in a potential state.
This is also corroborated by observations \citep[e.g.][]{hagyard90,hagyard:etal90,falconer:etal02}.
Similar observations also indicate that
it will usually be difficult to match the boundary magnetic field
with a linear force-free field. Also, from a more general point of view
it seems highly unlikely that the complex pre-eruption
coronal magnetic field which is
slowly but constantly
stressed by changing boundary conditions will have the current density
distribution of a linear force-free field, i.e. a constant $\alpha$.
A (strongly) localized current density in the erupting configuration seems
to be a lot more natural.
Another argument which is sometimes put forward to rule out linear force-free
magnetic fields for models of pre-eruptive states is that
linear force free fields
minimize the magnetic energy under the assumption
of global helicity conservation
\citep[e.g.,][]{taylor74}, and therefore if helicity conservation
could be assumed to hold for a coronal eruption, a linear
force-free field would not be able to provide the energy for
the eruption.
Even though it has been proposed to apply variants of
Taylor's hypothesis to the state of the coronal magnetic field
on larger scales
\citep[e.g.,][]{heyvaerts:priest84},
it generally does not seem to apply for models of eruptive processes
\citep[e.g.,][]{amari:luciani00}. Furthermore, there is the possibility
of helicity transport into and out of the region of interest and
for these reasons, we do not want to put too
much emphasis on this argument.

However, we believe that the other arguments are strong enough
to assume that generally one
has to take into account
that for the force free magnetic fields of pre-eruptive
states
the value of
$\alpha$ changes from field line to field line. This automatically
leads us to consider nonlinear force free fields.

\section{Extrapolation Methods}
\label{sec:methods}

The problem we discuss in
this section is to compute nonlinear force free magnetic fields
in Cartesian coordinates
if given all three components of the magnetic field on the boundary,
i.e. the photosphere. Various methods have been proposed
\citep[see e.g.][]{amari:etal97,demoulin:etal97,mcclymont:etal97,semel98,amari:etal00,yan:sakurai00}
of which we will discuss three. For the first method
we assume that the magnetic field is given on the lower boundary
(photosphere), whereas for the second and third method
we assume that ${\bf B}$ is prescribed on the surfaces of a
volume $V$.

\subsection{Direct Extrapolation}

Direct extrapolation \citep{wu:etal90} is a conceptually simple method, in which the Eqs.
(\ref{amperealpha}) and (\ref{alphaeq}) are reformulated in such way that
they can be used to extrapolate the photospheric boundary conditions given
by a vector magnetogram into the solar corona.

If we suppose that the boundary condition on the lower boundary
is given by ${\bf B}_0(x,y,0)$ then we can calculate the $z$-component of the current
density at $z=0$ by using Eq. (\ref{ampere})
\begin{equation}
\mu_0 j_{z0}=\frac{\partial B_{y0}}{\partial x}-\frac{\partial B_{x0}}{\partial y}.
\label{jz0direct}
\end{equation}
Knowing the $z$-component of the photospheric magnetic field ($B_{z0}$), we can
use Eqs. (\ref{jz0direct}) and (\ref{jparb}) to determine $\alpha(x,y,0)=\alpha_0$~:
\begin{equation}
\alpha_0=\frac{j_{z0}}{B_{z0}}
\label{alpha0direct}
\end{equation}
We remark that special care has to be taken at photospheric polarity inversion
lines, i.e. lines along which $B_{z0}=0$ \citep[see e.g.,][]{cuperman:etal91}.
Equation (\ref{alpha0direct})
allows us to calculate the $x$ and $y$-component of the current density
using Eq. (\ref{jparb})
\begin{eqnarray}
j_{x0 } & = & \alpha_0\, B_{x0},\\
j_{y0}  & = & \alpha_0\, B_{y0}.
\end{eqnarray}

We now use Eq. (\ref{solenoidal}) and the $x$ and $y$-component
of Eq. (\ref{ampere}) to obtain expressions for the $z$-derivatives of
all three magnetic field components in the form
\begin{eqnarray}
\frac{\partial B_{x0}}{\partial z} & = & j_{y0}+\frac{\partial B_{z0}}{\partial x}, \\
\frac{\partial B_{y0}}{\partial z} & = &\frac{\partial B_{z0}}{\partial y}-j_{x0},\\
\frac{\partial B_{z0}}{\partial z} & = & -\frac{\partial B_{x0}}{\partial x}-
\frac{\partial B_{y0}}{\partial y}.
\end{eqnarray}
The idea is to integrate this set of equations (numerically) in the direction of
increasing $z$, basically repeating the previous steps at each height.

Unfortunately it can be shown that the formulation of the force free equations
in this way is unstable because it is an ill-posed problem \citep[e.g.,][]{cuperman:etal90,amari:etal97}.
In particular one finds that
exponential growth of the magnetic field with increasing height is a typical
behaviour. The reason for this is that
the method transports information only from the photosphere upwards.
Other boundary conditions. e.g. at an upper boundary, either at a finite height or at infinity
cannot be taken into account.
Attempts have been made to regularize the method \citep[e.g.,][]{cuperman:etal91,demoulin:etal92}, but
cannot be considered as fully successful.

\subsection{MHD-Relaxation}

Another possibility to calculate nonlinear force free fields
is by the method of MHD relaxation \citep[e.g.,][]{chodura:schlueter81}. The idea is to start
with a suitable magnetic field which is not in equilibrium
and to relax it into a force free state. This is done by
using the MHD equations in the following form~:
\begin{eqnarray}
\nu  {\bf v }&=& (\nabla \times {\bf B}) \times {\bf B} \label{eqmotion}\\
{\bf E} + {\bf v} \times {\bf B} &=& {\bf 0 } \label{iohms}\\
\frac{\partial {\bf B}}{\partial t} &=&-\nabla \times {\bf E} \label{faraday} \\
\nabla \cdot {\bf B} &=& 0 \label{sole2}
\end{eqnarray}
The equation of motion (\ref{eqmotion}) has been modified in such a way
that it ensures that the (artificial) velocity field is reduced.
Equation (\ref{iohms}) ensures that the magnetic connectivity
remains unchanged during the relaxation.
The relaxation coefficient can be chosen in such way that it
accelerates the approach to the equilibrium state.
We use
\begin{equation}
\nu =  \frac{1}{\mu} \, |{\bf  B}|^2
\label{nudef}
\end{equation}
with
$\mu=$ constant. Choosing the relaxation coefficient $\nu$ proportional to
$B^2$ speeds up the relaxation process in regions of weak magnetic field
 \citep[see][]{roumeliotis96}.
Combining Eqs. (\ref{eqmotion}), (\ref{iohms}),
(\ref{faraday}) and (\ref{nudef}) we get an equation for the
evolution of the magnetic field during the
relaxation process~:
\begin{equation}
\frac{\partial {\bf B}}{\partial t} =\mu \; \nabla \times
\left(\frac{\left[(\nabla \times {\bf B}) \times {\bf B }\right]
\times {\bf B}}{B^2}  \right).
\label{relaxinduct}
\end{equation}
This equation is then solved numerically starting with a given
initial condition for ${\bf B}$. Equation (\ref{relaxinduct})
ensures that Eq. (\ref{sole2}) is satisfied during the
relaxation if the initial magnetic field satisfies it.

The difficulty with the relaxation method in the present form
is that it cannot be guaranteed that
for the boundary conditions we impose and
for a given initial magnetic field (i.e. given
connectivity), a smooth force-free equilibrium exists to which the
system can relax. If such a smooth equilibrium does not exist the
formation of current sheets is to be expected which will lead
to numerical difficulties. Therefore, care has to be taken when choosing
an initial magnetic field. This clearly limits the applicability
of the relaxation method, and further work will be needed to overcome this
obstacle. We therefore only apply the relaxation method for comparison
with the optimization method
to cases where we know the boundary conditions to be compatible.

One can show, however, that
the relaxation method converges under
less restrictive boundary conditions.
By multiplying Eq. (\ref{faraday}) by ${\bf B}$,
integrating over the complete computational volume $V$ and using
Eq. (\ref{iohms}), we find that
\begin{equation}
\frac{d}{d t} W_B =
 \oint_{\partial V}[({\bf v}\cdot{\bf B}) {\bf B} -
                    B^2 {\bf v}]\cdot {\bf n} dS
-\int_V ({\bf j}\times {\bf B})\cdot {\bf v} dV ,
\label{Benergy}
\end{equation}
where
\begin{equation}
W_B = \frac{1}{2}\int_V {\bf B}^2 dV .
\label{magenerg}
\end{equation}
Under line-tying boundary conditions we have
${\bf v}={\bf 0}$ on the boundary,
and only the volume integral survives. Using Eq. (\ref{eqmotion})
and the fact that $\nu >0 $ one can show that in this case
\begin{equation}
\frac{d}{d t} W_B = -\int_V \frac{\mu_0}{\nu}\; ({\bf j}\times {\bf B})^2 dV < 0 ,
\label{relaxconv}
\end{equation}
and an equilibrium is achieved if and only if the magnetic
field is force free.

We emphasize that these boundary conditions are usually not compatible
with prescribing all three magnetic field components on the boundaries,
but only the normal component. Therefore, in the following we discuss
the relaxation method only for test cases in which we know that
the boundary conditions are compatible. This is done for comparison
with the other method we discuss now, the optimization
method.
\subsection{Optimization}

Recently \citet{wheatland00} have proposed an optimization method
which refines a proposal by \citet{roumeliotis96}. In this
method the functional
\begin{equation}
L=\int_{V} \left[B^{-2} \, |(\nabla \times {\bf B}) \times {\bf B}|^2
+|\nabla \cdot {\bf B}|^2\right] \; d^3V
\label{defL}
\end{equation}
is minimized. Obviously, $L$ is bounded from below by $0$. This
bound is attained if the magnetic field satisfies the
force free equations
\begin{eqnarray}
(\nabla \times {\bf B}) \times {\bf B} &=& 0 \\
\nabla \cdot {\bf B} &=& 0.
\end{eqnarray}
The variation of $L$ with respect to an iteration parameter $t$ leads to
(see \citet{wheatland00} for details)
\begin{equation}
\frac{1}{2} \; \frac{d L}{d t}=-\int_{V} \frac{\partial {\bf B}}{\partial t} \cdot {\bf F} \; d^3V
-\int_{S} \frac{\partial {\bf B}}{\partial t} \cdot {\bf G} \; d^2S
\label{minimize1}
\end{equation}
where
\begin{eqnarray}
{\bf F} & =& \nabla \times ({\bf \Omega} \times {\bf B} )
- {\bf \Omega }\times (\nabla \times {\bf B})  \nonumber\\
& & -\nabla(\bf \Omega \cdot \bf B)+  \bf \Omega(\nabla \cdot { \bf B})
+\Omega^2 \;{ \bf B} \\
{\bf G} & = & {\bf \hat n} \times ({\bf \Omega} \times {\bf B} )
-{\bf \hat n} (\bf \Omega \cdot \bf B) \\
{\bf \Omega} & = & B^{-2} \;\left[(\nabla \times {\bf B})
    \times \bf B-(\nabla \cdot {\bf B}) \;{ \bf B} \right]
\end{eqnarray}
The surface term in (\ref{minimize1}) vanishes for
$\partial {\bf B}/\partial t=0$
\footnote{This condition makes it necessary that all three components of the magnetic field
have to be prescribed on the six boundaries of the computational box. In section
\ref{sec:randrelax} we discuss how the surface integral in (\ref{minimize1})
can be used to update the lateral and top boundary conditions during the
optimization process. Under this condition only the bottom boundary is
prescribed time independently with the  help of photospheric vector magnetograms.}
on the boundary and $L$ decreases
for
\begin{equation}
\frac{\partial B}{\partial t}=\mu \bf F
\label{iterate}
\end{equation}

This leads to an iteration procedure for the magnetic field which
is based on the equation
\begin{eqnarray}
\frac{\partial {\bf B}}{\partial t} & = & \mu \; \nabla \times
\left(\frac{\left[(\nabla \times \bf B) \times \bf B \right]
\times \bf B}{B^2}  \right) \nonumber\\
& &  + \;
\mu \;  \Bigl\{ \Bigl.
 - {\bf \Omega} \times (\nabla \times {\bf B}) -\nabla({\bf \Omega} \cdot {\bf B}) \nonumber\\
& & + \Bigl. {\bf \Omega}(\nabla \cdot {\bf B}) + \Omega^2 \; {\bf B}
\Bigl\}
\label{wheateq}
\end{eqnarray}
Here $\mu$ is a constant which can be chosen to speed up
the convergence of the iteration process.
Equation (\ref{relaxinduct}) and the leading terms
of Eq. (\ref{wheateq}) are identical, but Eq. (\ref{wheateq})
contains additional terms.

For this method the vector field ${\bf B}$ is not necessarily
so\-le\-noi\-dal during the computation, but will be divergence-free
if the optimal state with $L=0$ is reached. A disadvantage of the method
is that it cannot be guaranteed that this optimal state is indeed
reached for a given initial field and boundary conditions. If this is
not the case then the resulting ${\bf B}$ will either be not force free
or not solenoidal or both.

\section{Tests}
\label{sec:tests}

We encoded the MHD-relaxation method and the optimization method in one program.
The code is written in  {\it C} and uses 4th order finite differences on
an equal spaced grid. The time iteration is computed with the
method of steepest gradient \citep[see e.g. text books like][]{geiger}.
The program is parallelized with OpenMP.

\subsection{The semi-analytical test field}

To test the reconstruction methods, we try to reconstruct a
semi-analytic nonlinear force free solution found by \citet{lowandlou}.
\citet{wheatland00} have used similar tests. The main difference between their paper and ours
is in the diagnostic quantities used and in the comparison with i) the relaxation method and ii)
the noisy boundary data (which we think is more representative of a realistic situation).
\citet{lowandlou} solved the Grad-Shafranov equation for axis-symmetric
force free fields
in spherical coordinates $r$, $\theta$, $\phi$.
For axis-symmetry the magnetic field can be written in the form
\begin{equation}
{\bf B} = \frac{1}{r \sin\theta} \; \left(
\frac{1}{r} \, \frac{\partial A}{\partial \theta} {\bf e}_r-
\frac{\partial A}{\partial r} {\bf e}_{\theta} +
Q \, {\bf e}_{\phi} \right)
\label{lowloub}
\end{equation}
where $A$ is the flux function, and $Q$ represents the $\phi$-component
of ${\bf B}$, depending only on $A$.
The flux function $A$ satisfies the Grad-Shafranov equation
\begin{equation}
 \frac{\partial^2 A}{\partial r^2}+\frac{1-\mu^2}{r^2} \,
\frac{\partial^2 A}{\partial \mu^2}+Q \; \frac{d \, Q}{d \, A} =0
\end{equation}
where $\mu=\cos\theta$.
Among others, \citet{lowandlou} derive solutions for
\begin{equation}
\frac{d  Q}{d  A} = \alpha= \mbox{constant}
\end{equation}
by looking for separable solutions of the form
\begin{equation}
A(r,\theta)=\frac{P(\mu)}{r^n}.
\end{equation}
The solutions are axis-symmetric in
spherical coordinates with a point source
at the origin, but if used for testing force free codes
in Cartesian geometry the symmetry
is no longer obvious after a translation which places
the point source outside the computational domain
and a rotation of the symmetry axis with respect to
the Cartesian coordinate axis.

\subsection{Test runs}
\label{sec:app}

For the test we have first calculated
$B_z(x,y,0)$ on the photospheric boundary $z=0$ for the
appropriate Low and Lou solution. Using this $B_z$ as
a boundary condition we
calculate the potential field corresponding to this
boundary conditions inside our computational box.
The potential field is computed with help of a method developed
by \cite{seehafer78}. This method gives the components of the magnetic
field in terms of a Fourier series. The observed magnetogram
(or here the extracted magnetogram from the Low and Lou solution)
covers a rectangular region extending from $0$ to $L_x$ in $x$ and $0$ to $L_y$
in $y$ is artificially extended onto a rectangular region covering $-L_x$ to $L_x$
and $-L_y$ to $L_y$ by taking an antisymmetric mirror image of the original
magnetogram in the extended region, i.e.
$B_z(-x,y) = -B_z(x,y)$ and $B_z(x,-y) = -B_z(x,y)$.
We use a Fast Fourier Transformation
(FFT) scheme (see  \cite{alissandrakis81}) to
determine the coefficients of the Fourier series.
For more details regarding this method see \cite{seehafer78,seehafer82}.

The potential field is used as starting
field inside
the computational box but
the Low and Lou ${\bf B}$ field is imposed on the boundaries.
We then use either the MHD relaxation or the
optimization method to calculate the correct force free
equilibrium field. During the computations we calculate
the quantities $L/[T^2 m]$ (for both relaxation and
the optimization method), the absolute
value of the Lorentz force $|{\bf J} \times {\bf B}|/[nN \; m^{-3}]$
(averaged over the numerical grid), the value
of $|\nabla \cdot{\bf B}|/[\mbox{mGauss} \, m^{-1}]$ (averaged over the numerical grid),
and the difference
between the numerical magnetic field and the
known analytical solution
$\frac{|{\bf B}(t)- {\bf B}_{ana}|^2}{|{\bf B}_{ana}|^2}  $
(averaged over the numerical grid)
at each time step.
In section \ref{sec:standard} and \ref{sec:noise} all components of the
magnetic field are fixed on all six boundaries. In section \ref{sec:randrelax}
only the bottom boundary condition is prescribed time-independent and the
lateral and top boundaries are updated during the iteration. The details are
described in \ref{sec:randrelax}.
\begin{figure*}
\begin{center}
\figbox*{}{}{%
\includegraphics[width=6.0cm]{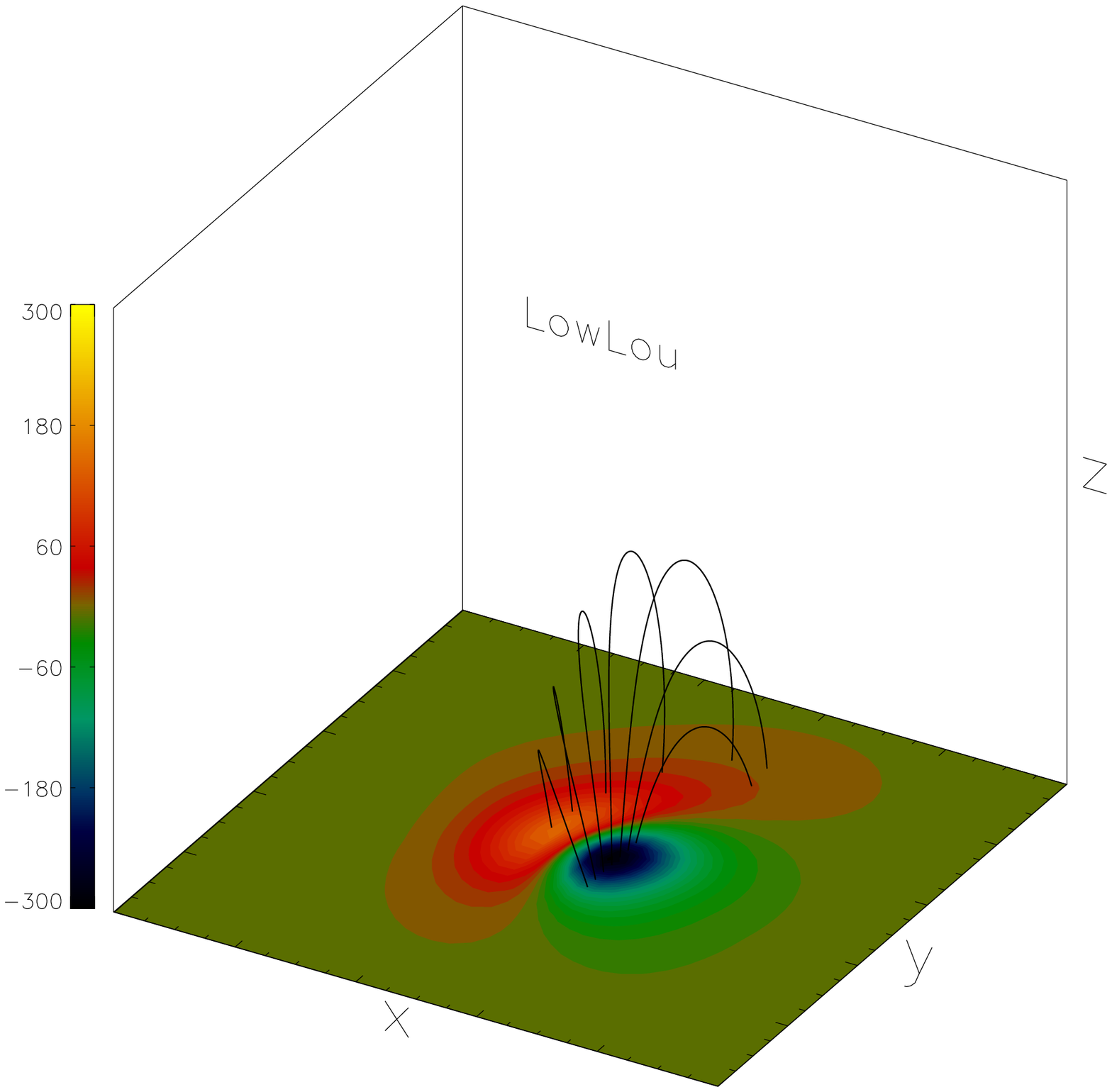}
\includegraphics[width=6.0cm]{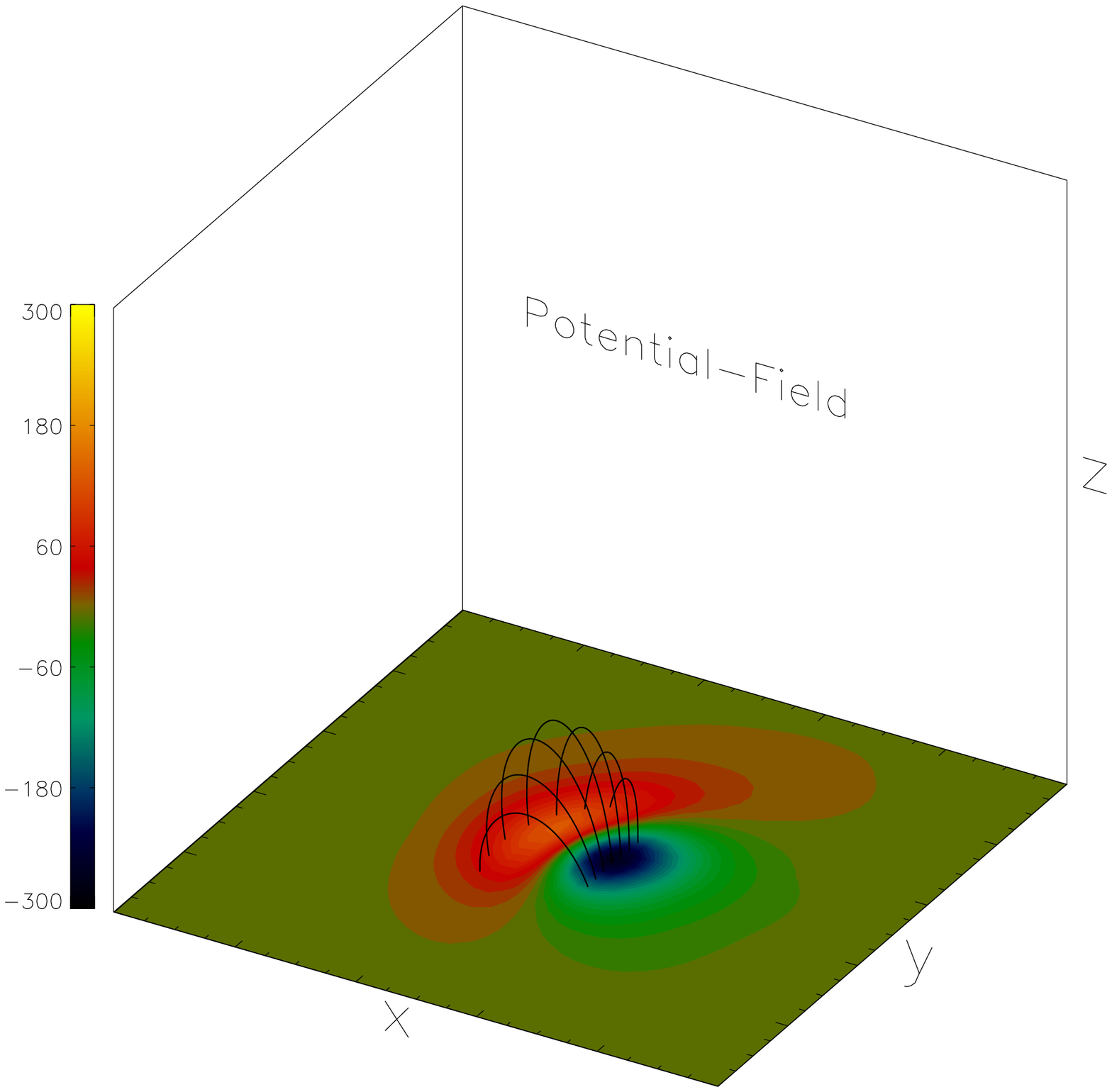}}
\figbox*{}{}{%
\includegraphics[width=6.0cm]{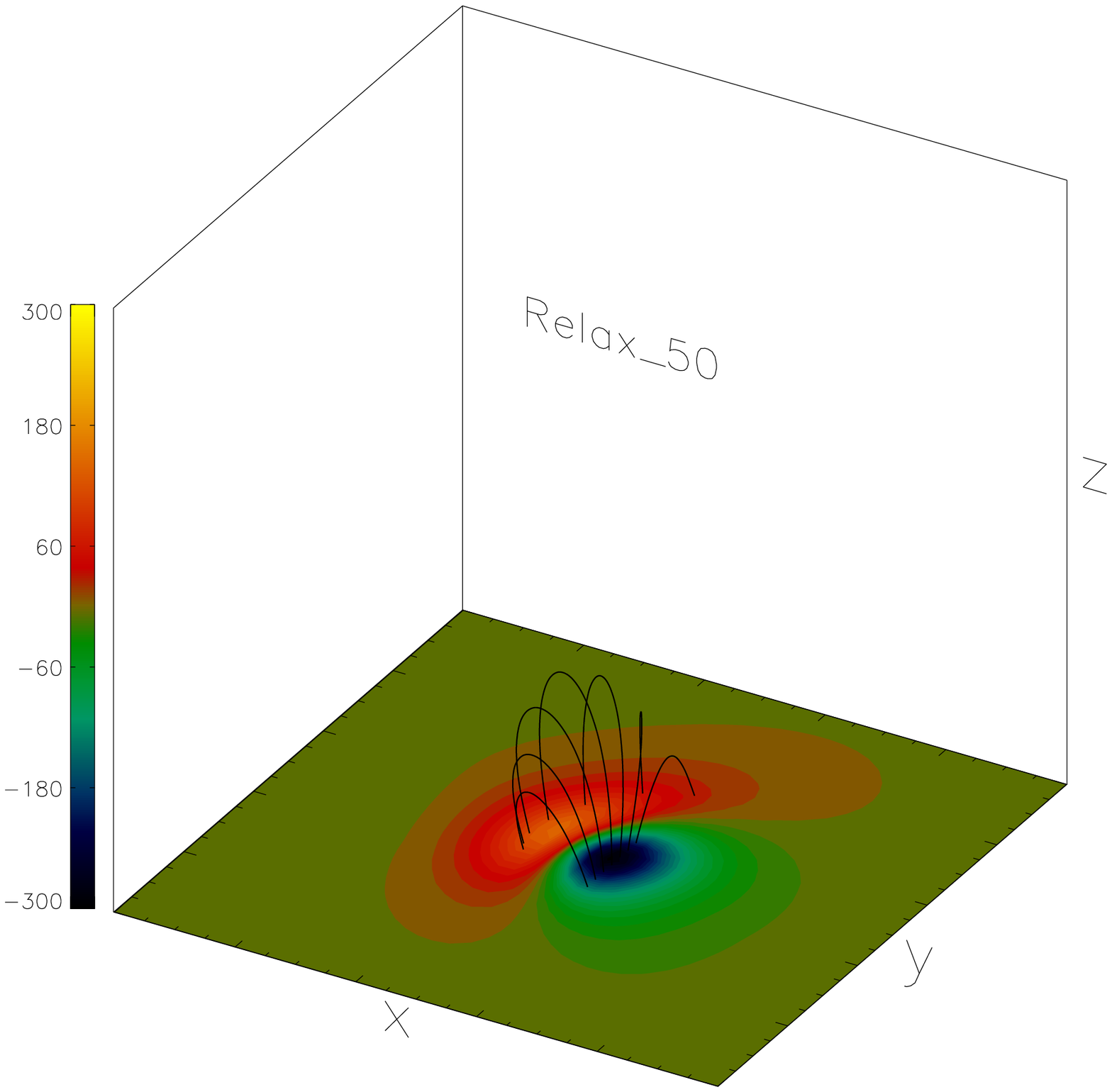}
\includegraphics[width=6.0cm]{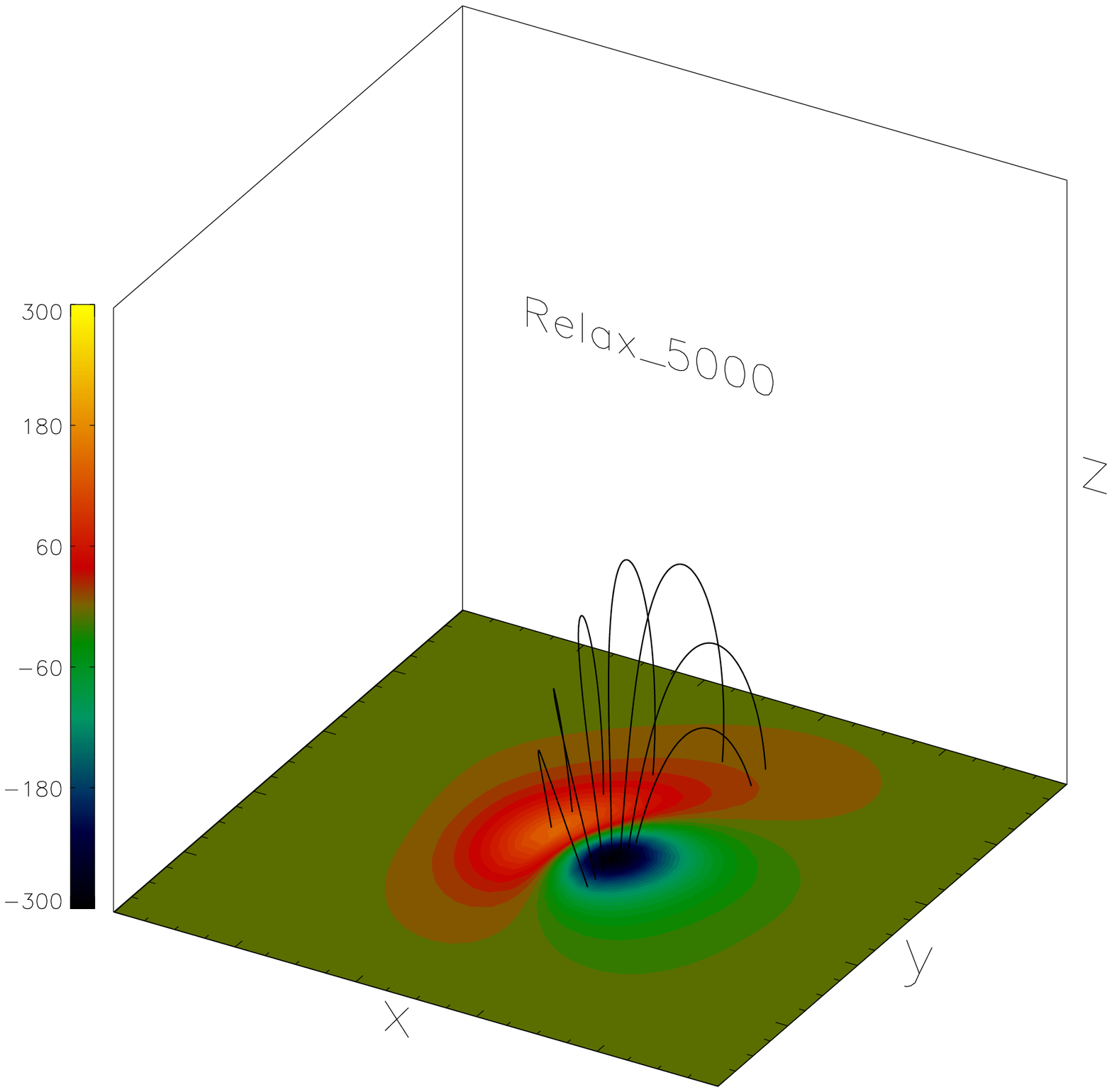}}
\figbox*{}{}{%
\includegraphics[width=6.0cm]{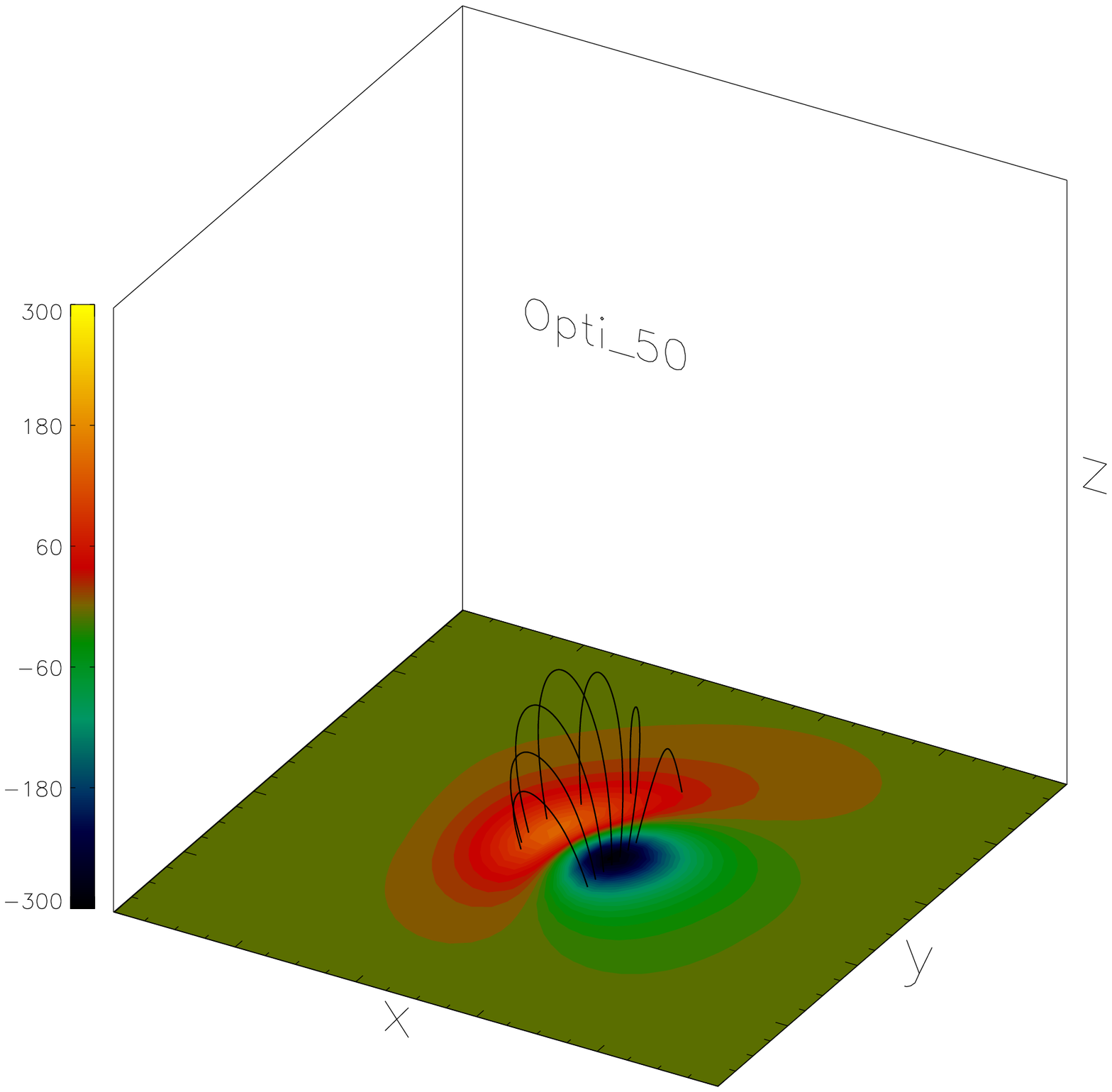}
\includegraphics[width=6.0cm]{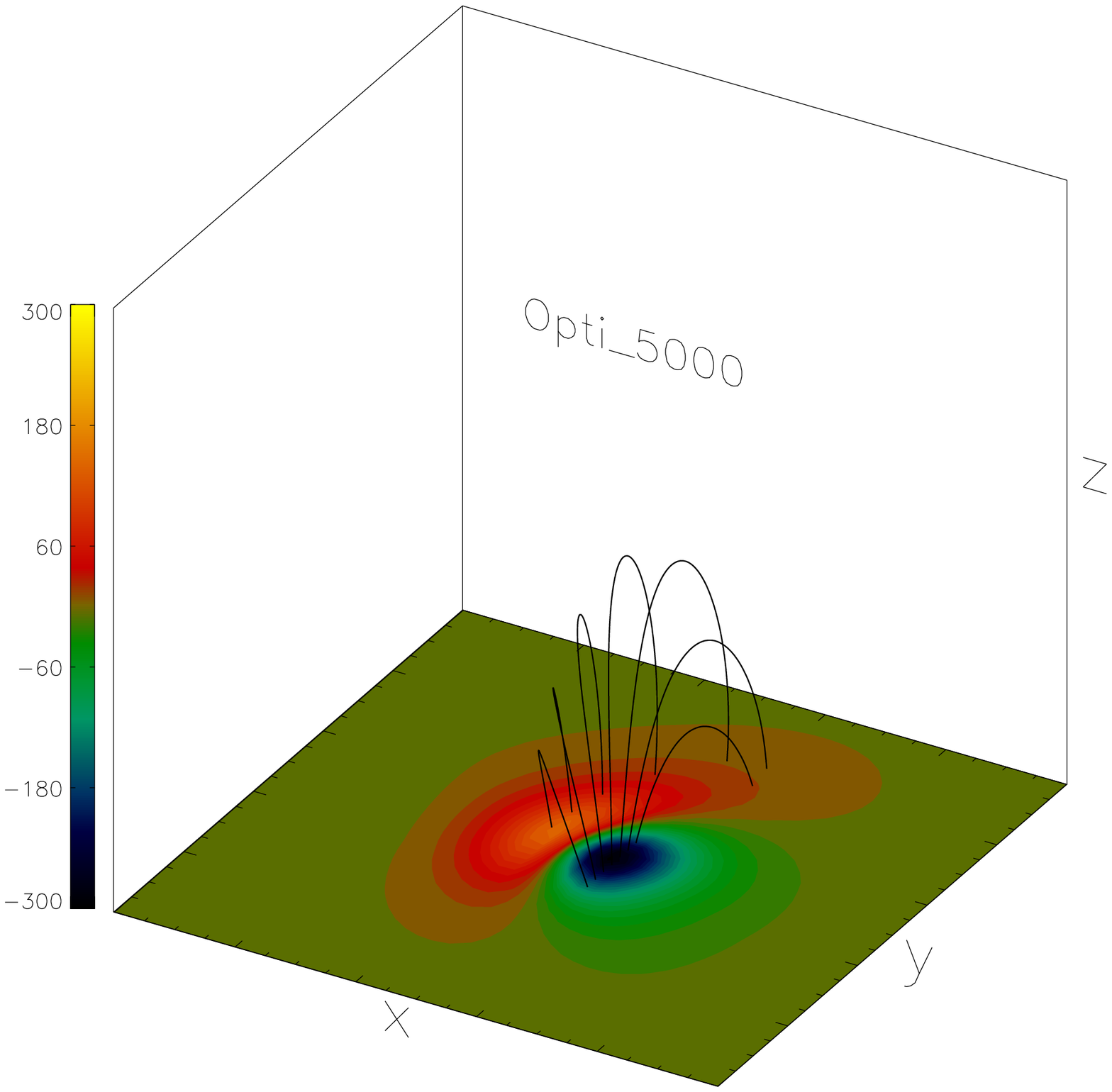}}
\end{center}
\caption{Top row, left panel: A set of selected field lines of the Low and Lou solution.
Top row, right panel: The same field lines for the corresponding potential field
calculated as described in the main text. The difference between the two fields is obvious.
Middle row: The field lines after $50$ (left) and 5,000 (right) steps of the MHD relaxation method.
Bottom row: The field lines after $50$ (left) and 5,000 (right) steps of the optimization method.
The box drawn shows the spatial extension of the numerical box.
The colour coding at the bottom of the boxes shows the normal component of the photospheric
magnetic field in Gauss.}
\label{3dlines}
\end{figure*}
\subsubsection{Standard tests}
\label{sec:standard}
In Fig. \ref{3dlines} we show three-dimensional plots of
selected magnetic field lines for the Low and Lou solution,
the potential field calculated by taking the Low and Lou $B_z$
on the lower boundary, the field of the MHD relaxation
method after $50$, $500$ and $5,000$ relaxation time steps, and
the same plots for the optimization method. In these runs we used
a grid size of $40 \times 40 \times 20$. The colour coding
of the bottom boundary indicates the $B_z$ distribution on that
boundary.

It can bee seen that the potential field which is used as starting
field of the computations for both methods is clearly different
from the
Low and Lou test field. The state of the system after $50$ steps
still shows some resemblance to the initial potential field
for both methods but the field lines have started to evolve away from
the potential field. One can also see small differences between
the two methods. After $5,000$ steps no obvious differences between
the magnetic field reached with either method and the Low and Lou field
can be seen.

\begin{figure}
\includegraphics[width=8.0cm,height=16.0cm]{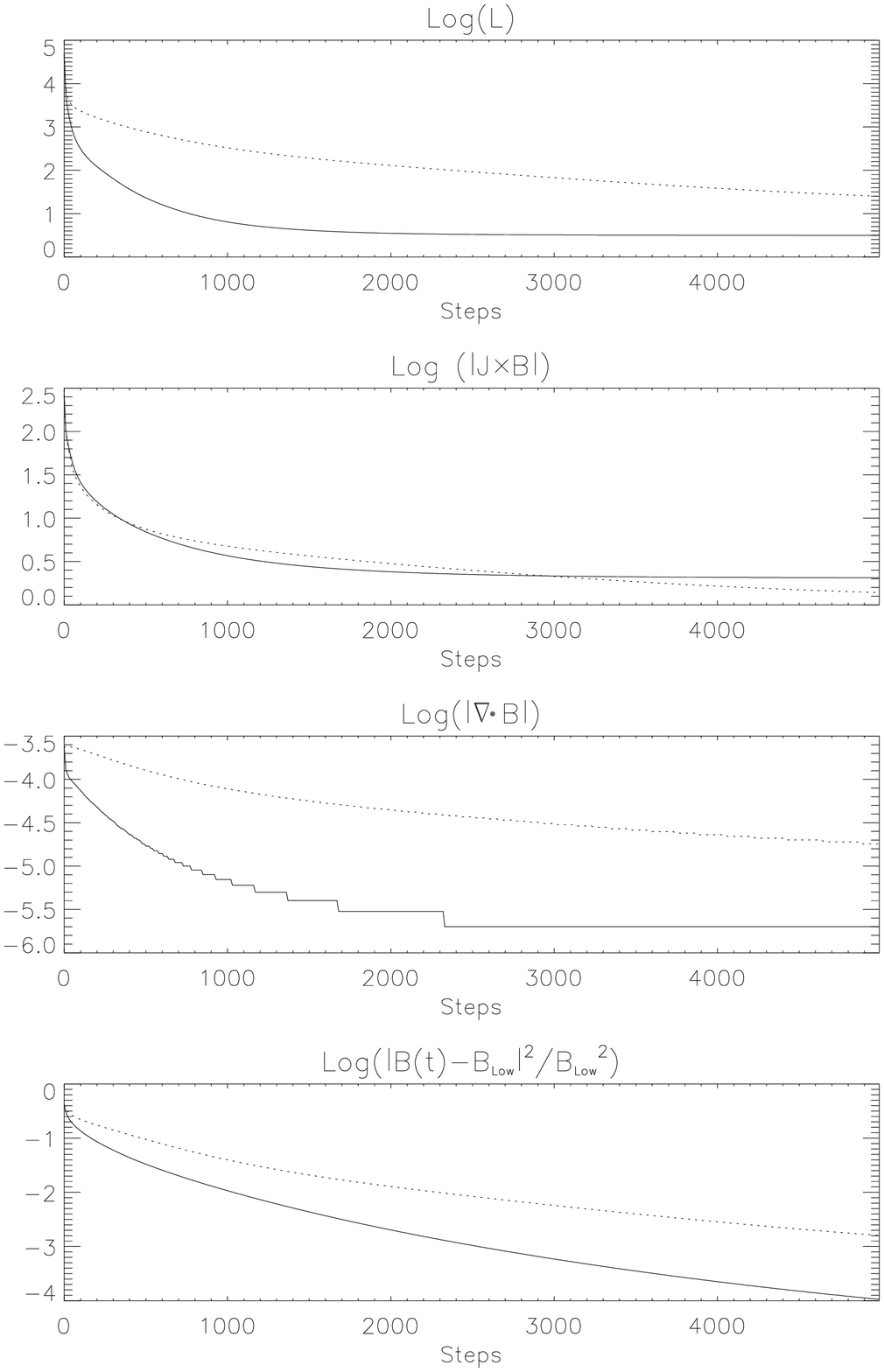}
\caption{Evolution of the diagnostic quantities.
Top panel: functional $L$ (defined in Eq. (\ref{defL})),
second panel: Lorentz force $|{\bf j} \times {\bf B}|$ (averaged over the numerical grid),
third panel: $|\nabla \cdot {\bf B}|$ (averaged over the numerical grid)
bottom panel: relative error $\frac{|{\bf B}(t)- {\bf B}_{ana}|^2}{|{\bf B}_{ana}|^2}  $
(averaged over the numerical grid).
The quantities are drawn for the  MHD relaxation method
(dotted lines) and for the optimization method (solid lines).
The grid size used was $40 \times 40 \times 20$.}
\label{diagnostic1}
\end{figure}

To quantify this statement we show in Fig. \ref{diagnostic1} a comparison
of the evolution of the four diagnostic quantities for the two methods
during the computation. For both methods all four diagnostic quantities
decrease during the computation but after $5,000$ steps the optimization
method shows significant smaller values for the quantities $L$, $\nabla \cdot {\bf B}$
and the comparison with the analytic solution.
Therefore we can state
that in this case the optimization method seems to be slightly more
efficient than the MHD relaxation method.
\begin{table}
\caption{Details of runs to reconstruct a \protect\citet{lowandlou}
solution.  The values of the parameters used by Low and Lou are $l=0.3$ and $ \Phi=\pi/4$.
The first column contains the used grid size and comments which code (relaxation or optimization)
and boundary conditions have  been used. If not specified the boundary
conditions have been extracted from the analytic solution. We specify  if
noise is added to the boundary conditions. The remark {\it Pot-boundary}
means that the lateral and top boundary conditions are not extracted from the
analytic solution, but prescribed as a global potential field.
The second column contains the iteration step.
The third column contains the value of the functional $L$, the fourth column
the Lorentz force (averaged over the numerical grid) and the last column the
relative error compared with the analytic solution.}
\begin{tabular}{|r|r|r|r|r|r|}
\hline
$n_x \times n_y \times n_z$ &Step & $\frac{L}{[T^2 m]}$ & $\frac{|{\bf j} \times {\bf B}|}{[nN \, m^{-3}]}$& Relative Error \\
\hline
$40 \times 40 \times 20 $ &{\bf discret. error}&$ 3.2$&$ 2.0 $ & Reference \\
Start & 0&$46573 $&$264$ &$0.42$ \\
\hline
Relaxation & 50 &$3092$&$38.2$ &$0.26$  \\
$ $ & 500 &$769 $&$7.5$ &$0.095$ \\
$ $ &5,000 &$ 25.0 $&$ 1.4$ &$0.0016$ \\
\hline
Optimization & 50 &$832$&$44.1$ &$0.19$   \\
$ $ & 500 &$23.1 $&$7.0$ &$0.03$ \\
$ $ &5,000 &$ 3.15 $&$ 2.05$ &$0.0001$ \\
1\% noise &5,000&$9.5$&$3.7$ &$0.00017$ \\
10\% noise &5,000&$645.0$&$29.2$ &$0.0055$ \\
20\% noise &5,000&$2584$&$58.6$ &$0.028$ \\
Pot-boundary &500&$468$&$9.1$ &$0.39$ \\
Pot-boundary &5,000&$458$&$6.3$ &$0.37$ \\
 \hline
 \hline
$80 \times 80 \times 40 $ &{\bf discret. error}&$ 0.6$&$0.65$ &Reference  \\
Start & 0&$109203 $&$297$ &$0.48$  \\
\hline
Optimization &50 &$1590$&$52$ &$0.26$  \\
$ $ &500& $81$ &$12$ &$0.079$  \\
$ $ &5,000&$0.79$&$1.02$ &$0.0013$  \\
$ $ &25,000&$0.6$&$0.65$ &$0.000003$  \\
10\% noise &25,000&$1876$ &$37.6$&$0.0078$  \\
Pot-boundary &1,000&$563$&$12.8$ &$0.42$  \\
Pot-boundary &10,000&$529$&$5.3$ &$0.36$ \\
 \hline
\end{tabular}
\label{resultstab}
\end{table}
This is corroborated by a look a Table \ref{resultstab} in which we summarize
the main results of the various test runs
we have made. The first row shows the discretisation errors for $L$ and the Lorentz force
if the known solution is discretised on
a $40\times 40\times 20$-grid and  used to calculate the values
of this quantities. The second row contains the values of $L$, the Lorentz force
and the relative error after the interior grid points have been replaced by the
potential field calculated from the photospheric $B_z$ of the Low and Lou solution.
The relatively large values of the three diagnostic quantities show the deviation
from the equilibrium.

The next two rows show how the three diagnostic quantities evolve during the
MHD relaxation method for the $40\times 40\times 20$-grid.
One can see that after $5,000$ steps the value of $L$ has dropped
by almost three orders of magnitude, but is still  one order of magnitude
above the value calculated with the discretised exact solution. The Lorentz force has
dropped to the level of the dicretisation error, and the relative error
has dropped more than two orders of magnitude.

We are giving the same quantities for the optimization method in the following three
rows. It can be seen that especially the values of $L$ are always way below the corresponding
values of the MHD relaxation method. This is not surprising as the optimization method
relies on the minimization of the functional $L$. The final value of $L$ is actually below
the discretisation error.
It can also be seen that  the relative error after $5,000$ iterations is
more than one order of magnitude smaller than the corresponding
value of the MHD relaxation method.

We have made a step towards checking numerical convergence for the optimization method by
repeating the calculation on a grid with doubled resolution ($80\times 80 \times 40$).
The results are shown in the lower part of Table \ref{resultstab}.
The first thing to notice is that the discretisation error is almost an order of
magnitude smaller than
for the previous grid.
For this grid
$25,000$ iteration steps have been carried out, and after that the values of $L$ and
the Lorentz force have reached the level of the discretisation error. The
relative error is more than two orders of magnitude smaller than for the coarser grid.
\subsubsection{Effect of adding noise}
\label{sec:noise}
The previous calculations have been carried out under the assumption that the magnetic
field on the boundary of the computational box is known exactly. Such an idealized situation
will not be found when real data are used. Vector magnetograms
will have finite resolution and suffer from observational uncertainties
making the reconstruction potentially more difficult. Especially the
optimization method has been proposed in view of coping with these
difficulties \citep{wheatland00}.

\begin{figure}[t]
\figbox*{}{}{%
\includegraphics[width=8.0cm,height=8.0cm]{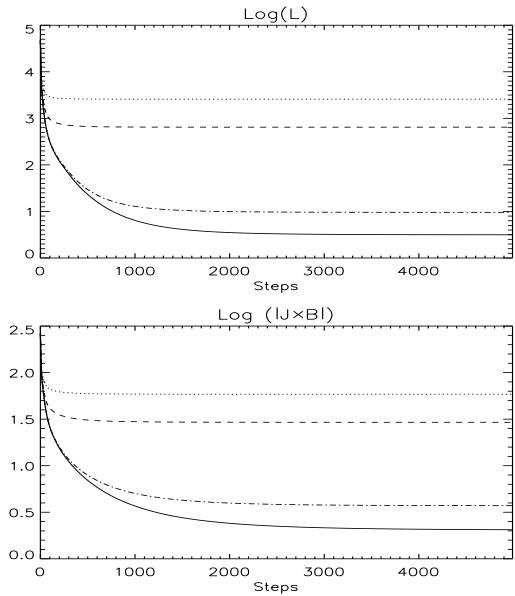}
}
\caption{Evolution of $L$ and $|{\bf j} \times {\bf B}|$ for the
optimization code
for various levels of boundary value noise.
The lowest lines correspond to an ideal vector magnetogram and the higher lines
to vector magnetograms with $1\%$, $10\%$ and $20\%$ noise, respectively. The grid
size used here is $40 \times 40 \times 20$. The convergence of the method gets smaller
with increasing noise level.}
\label{noisefig}
\end{figure}
To investigate how the
optimization method works if the boundary conditions
are not given by the exact analytical field, but to
keep control over the amount of uncertainty, we have carried
out test runs with the optimization method adding random noise with
to the boundary conditions. We add the noise by multiplying the exact
boundary conditions with a number $1+\delta$ where $\delta$ is
a random number in the range $-n_l \leq \delta \leq n_l$
and $n_l$ is the noise level.

To study the effect of the noise we have done runs
with different amplitudes of noise on the $40 \times 40 \times 20 $ grid.
The evolution of $L$ and the Lorentz force with the numbers of iteration
for various noise levels are shown in Fig. \ref{noisefig}. It can clearly
be seen that the method converges less and less well with increasing
noise. For noise levels of 10 \% and 20 \%, the corresponding values
after $5,000$ steps for the
$40 \times 40 \times 20$ grid are given in Table \ref{resultstab}. We
have also carried out a run on the $80 \times 80 \times 40$ grid
with a noise level of 10 \%. For this run the values of the
diagnostic quantities after $25,000$ steps are still higher than the values
of the corresponding quantities on the coarser grid after $5,000$ steps.
We have to conclude that even a 10 \% uncertainty in the boundary conditions
could affect the convergence of the optimization method quite badly.
The main problem is that with noise and fixed boundary conditions
on all six boundaries the boundary condition are over-imposed and will no longer
be compatible. We discuss how this problem can be solved in the next section.
\subsubsection{The problem of the lateral and top boundary conditions}
\label{sec:randrelax}
Until now the runs have been carried out under the assumption that the
magnetic field boundary conditions are well known on all six boundaries of
the box. Unfortunately  real vectormagnetograms only provide information
regarding the photospheric magnetic field. The top and side
boundary conditions are unknown and have to be fixed somehow.
Here we want to investigate the influence of the choice of these
boundary conditions. An additional problem  occurs by fixing the vector
magnetic field on all six
boundaries (as done in the previous sections) because one overimposes the boundary
conditions. For the test cases in the previous sections the boundary
conditions have been extracted from an analytic solution and consequently the
boundaries are automatically compatible (this implies that there are no
problems with over-imposed boundary conditions). Section \ref{sec:noise}
showed that this compatibility might get lost if the boundaries contain
noise. To overcome this difficulty and to take into account that measured data are
only available for the bottom boundary we describe how the strict condition
$\frac{\partial {\bf B}}{\partial t}=0$ on all boundaries  can be avoided within the
optimization procedure. First we have to impose the boundary conditions as a
well posed problem. A popular well posed boundary condition for non linear
force free fields is to impose the normal component $B_n$ of the magnetic field and
the normal component of the current density $j_n$ in regions for a positive (or negative)
$B_n$.
\citep[see][ regarding the compatibility of
photospheric vector magnetograph data]{aly88}. These conditions
have been derived
under the very strict constraint of a flux balanced magnetogram
 where all magnetic flux
is closed.
This implies that each magnetic field line starting at one point
on the photosphere also has to end on the photosphere (in a region of
opposite magnetic flux, but with the same value of $\alpha$). Such strictly
isolated active regions seem to be a too restrictive constraint for real vector
magnetograms.
Real vectormagnetograms provide all three components of the
magnetic field for either sign of $B_n$ and thus also $j_n$ on the complete
bottom boundary. It seems reasonable to impose these observed data on the bottom
boundary. The price we have to pay is that field lines starting on the
photosphere might pass the lateral and top boundary.
\footnote{If the flux is not balanced in the region, it implies that
part of the flux distribution is missing from the limited-size of the
magnetogram or that the calibration of the magnetograph is bad. In both
cases, any magnetic extrapolation method will only derive an approximate
field. Unfortunately the size of observed vectormagnetograms is limited and
a real magnetogram will usually not be exactly flux balanced. Field lines passing
the boundary of the computational box can be either open field lines or close
on the photosphere outside the observed magnetogram.}
Consequently we need
to update the lateral and top boundary during the iteration.

As a first step towards consistent boundary conditions for the
optimization method, we
choose a potential field on the lateral and top boundaries.
Firstly, a potential field can be easily computed just from
the measured normal component of the photospheric magnetic field and secondly,
it can be assumed that the solar magnetic field
is reasonably well approximated by a potential
field outside active regions. Here we compute the potential field on the
boundary as a global potential field computed from the $B_z$ distribution
at $z=0$ alone.
Table \ref{resultstab} shows that the value of
$L$ for the $40 \times 40 \times 20$ grid drops to a slightly lower value than
for the corresponding run with $10 \%$ noise, while the remaining
$|{\bf j}\times {\bf B}|$ forces are a factor of three lower. The relative error
compared with the exact solution is quite high here, which is no wonder
because the magnetic field is forced to stay potential close to the side
and top
boundaries. One might notice that the error in the forces is a factor of $3$
larger than the discretisation error, while $L$ is more than two
orders of magnitude larger
than the discretisation error. The reason is that an inconsistency occurs
at the edges of the box
between the photospheric boundary and the side boundaries. The Low and
Lou solution is not potential on the photosphere close to the side
boundaries, but the chosen side boundary conditions are potential.

To overcome this difficulty we cannot fix the lateral and top boundaries
during the iteration, but have to relax also these boundaries.
As the iteration equation (\ref{iterate}) has been derived under the
condition $\frac{\partial {\bf B}}{\partial t}=0$ on all boundaries
we have to extend the optimization approach. If we allow
$\frac{\partial {\bf B}}{\partial t} \not=0$ on some boundaries the
surface term in equation (\ref{minimize1}) does not necessarily vanish.
It is straightforward to extend the iteration by
\begin{equation}
\frac{\partial B}{\partial t}=\mu \bf G
\label{updateboundary}
\end{equation}
on the open boundaries. Equation (\ref{updateboundary}) changes the boundary values
in such way that $L$ decreases. In the following we choose the lateral
and top boundaries as open (they are initialized with a potential field)
and apply Eq. (\ref{updateboundary}) here. The bottom boundary remains time
independent during the iteration. Let us remark that one might as well
use the above described photospheric boundary condition for isolated active
regions ($j_n$ fixed on the photosphere only in regions of positive $B_n$).
To do so one has to apply Eq. (\ref{updateboundary}) also for the transversal
magnetic field on the photosphere in regions with negative $B_n$.

If we do not keep
the lateral and top boundary conditions fixed,
but allow them to relax with help of Eq. (\ref{updateboundary}),
 L drops to $36.2$ which is only one order of magnitude above the
discretisation error. The corresponding value of the total force decreases
slightly ($|{\bf j} \times {\bf B} |=5.3$) compared to fixed boundary conditions.
For observed vectormagnetograms it might be useful to choose a
sufficiently large area around
an active region, i.e. with the side boundaries relatively  far away from the
non-potential active region,
so that the magnetic field can be assumed to be approximately potential
close to
the side boundaries. We remark that it might be hard to find
corresponding vector magnetograms as these instruments (e.g. IVM in Hawaii)
do not observe the complete solar disk, but only restricted areas.
We have also carried out a run on the $80 \times 80 \times 40$ grid
with potential field boundary conditions. For this run
a stationary state is
reached after $10,000$ steps, and
the forces are less than one order of magnitude higher than the discretisation
error, while $L$ is nearly three orders of magnitude above the discretisation
error. Please note that the absolute error in the $|{\bf j} \times {\bf B}| $
has the same order of magnitude here as for the corresponding
$40 \times 40 \times 20$ run. If we allow relaxation on the side and top
boundaries, L drops to $96$ and $|{\bf j} \times {\bf B}|=8.7$ is slightly
higher than for fixed side boundary conditions.
We conclude that the uncertainty of the side and
top boundary conditions affects the
convergence of the optimization method. Its
influence can be reduced significantly if we allow for a relaxation of these
boundaries and only fix the photospheric boundary conditions, but further
improvements
would still be welcome.

\section{Conclusions}
\label{sec:end}

In the present paper we have presented an assessment of the properties
of an optimization method which has recently
been proposed for the calculation of nonlinear
force-free fields from given boundary data \citep{wheatland00}.
One part of the assessment was a comparison the
performance of the optimization method
with the performance of an MHD relaxation method. For both
methods new parallelized codes have been developed.
Both methods have been applied to finding a
known semi-analytic solution \citep{lowandlou} from a
given non-equilibrium initial condition.
Both methods converge to the exact solution, but
the optimization method has higher accuracy. The
MHD relaxation method is only applied to this case
since in its present numerical implementation, the boundary
conditions could give rise to inconsistencies.

To simulate a more realistic situation which is
closer to working with observational data from vector magnetograms
we added noise to the boundary conditions. We have only applied the
optimization method to this type of problem and found that already
a relatively
small noise level of 10 \%
can affect the convergence of the method
quite considerably. More work is needed to see how this difficulty
can be overcome. We also intend to generalize the
relaxation method in such way that is able of coping with
more realistic boundary data.

The ultimate aim for the future is to apply
these methods to data from vector magnetograms.
Unfortunately vector magnetograms are less accurate than line-of-sight
magnetograms (e.g. from MDI on SOHO). Therefore, in the light of our
results in the cases where noise was added to the boundary conditions
it will be interesting to see whether and how quickly the methods converge.
We also plan to use stereoscopic information as a further constraint
in the reconstruction process. This  will become especially important
for the analysis of data from the STEREO mission.
A method which is based on
linear force free fields has recently been proposed by \citet{twtn02},
but linear force free fields seem to be too restrictive to describe
coronal phenomena appropriately. Therefore, the natural
next step will be the use of nonlinear force free fields in such
a code.

\balance 



\begin{acknowledgements}
The authors thank Bernd Inhester and Hardi Peter for useful discussions.
TW acknowledges support by the European Communitys Human Potential Programme
through a Marie-Curie-Fellowship. TN thanks PPARC for support by an
Advanced Fellowship. The computations have been carried out on
the PPARC funded MHD Cluster in St. Andrews. We thank Norbert Seehafer and an
unknown referee fore useful comments.
\end{acknowledgements}


\end{document}